\documentclass[final,5p,times,twocolumn]{elsarticle}

\biboptions{sort&compress}

\usepackage[utf8]{inputenc}
\usepackage[T1]{fontenc}
\usepackage{amsmath,amssymb,amsfonts,mathtools}
\usepackage{graphicx}
\usepackage{bm}
\usepackage[
  colorlinks=true,
  linkcolor=blue,
  citecolor=blue,
  urlcolor=blue
]{hyperref}

\journal{Physics Letters B}

\newcommand{\dd}{\mathrm d}
\newcommand{\Kpt}{K_{\rm PT}}
\newcommand{\Kk}{K_K}

\newcommand{\sech}{\operatorname{sech}}

\begin{document}
\raggedbottom

\begin{frontmatter}

\title{Localised zero modes enhance massive Casimir interactions}

\author[uva,imuva]{Jose M. Mu\~noz-Casta\~neda\corref{cor1}}
\ead{jose.munoz.castaneda@uva.es}
\author[unizar,capa]{In\'es Cavero-Pel\'aez}
\author[uva]{Gonzalo Sancho-Garrido}

\cortext[cor1]{Corresponding author}

\address[uva]{Departamento de F\'isica Te\'orica, At\'omica y
\'Optica, Universidad de Valladolid, 47011 Valladolid, Spain}

\address[imuva]{IMUVa, Universidad de Valladolid,
47011 Valladolid, Spain}

\address[unizar]{Departamento de F\'isica Te\'orica,
Universidad de Zaragoza, 50009 Zaragoza, Spain}

\address[capa]{Centro de Astropart\'iculas y F\'isica de Altas
Energ\'ias, Universidad de Zaragoza, 50009 Zaragoza, Spain}
\begin{abstract}
Massive quantum fields ordinarily produce strongly screened Casimir
interactions at large separation. We show that a normalisable localised
zero mode can change this asymptotic behaviour: it opens an additional
interaction channel whose decay is parametrically slower than that due to
the exchange of massive quantum fluctuations, even though the continuum
itself remains gapped. We demonstrate this mechanism analytically and show that the massive quantum
fluctuations provide the local dressing of each object's coupling to the
zero-mode fluctuations. 
The exactly solvable Pöschl–Teller fluctuation operator of the sine-Gordon kink provides an explicit realisation: the one-loop Casimir interaction exhibits the predicted zero-mode-dominated large-distance behaviour, with the enhancement localised around the kink.

\end{abstract}


\end{frontmatter}

\section{Introduction}
\label{sec:introduction}
%%%%%%%%%%%%%%%%%%%%%%%%%%%%%%%%%%%%%%%%%

Quantum vacuum fluctuations generate interactions between neutral
objects, giving rise to the Casimir effect and its generalisations
\cite{Casimir1948,Milton2001,Bordag2009}. For massless quantum fields,
the Casimir interaction decays as an inverse power of the separation,
with an exponent fixed by the dimensionality and geometry of the
problem. If the fluctuating field has mass \(m>0\), the large-distance
interaction is instead exponentially screened. For two parallel plates
in \(D\) spatial dimensions,
\begin{equation}
\frac{E_0(d)}{S_{D-1}}
\sim \frac{C_0}{d^D},
\qquad
\frac{E_m(d)}{S_{D-1}}
\sim C_m\left(\frac{m}{d}\right)^{D/2}e^{-2md},
\quad md\gg1 ,
\end{equation}
where $d$ is the distance between plates and \(S_{D-1}\) denotes the area of the plates. Thus Casimir
interactions mediated by quantum fluctuations associated with massive
particles become rapidly irrelevant at distances large compared with
the Compton wavelength.

An inhomogeneous background can alter this infrared picture by
supporting discrete modes below the massive continuum. Among them, an
isolated normalisable zero mode is qualitatively distinct from the
remaining gapped spectrum and can open an additional infrared channel
whose large-distance decay is not governed by the direct exchange of
massive quantum fluctuations. Vacuum effects in solitonic and defect
backgrounds have been studied in a variety of settings
\cite{Rajaraman1982,Coleman1985,CaveroGuilarte2010,BordagMunoz2012,Fosco2016,GrahamWeigel2022}.
More generally, fluctuation-induced forces are sensitive to soft or zero-energy sectors, including critical and pseudo-Casimir modes
\cite{Gambassi2009,Ajdari1992PseudoCasimir}, Majorana zero modes
\cite{Beenakker2024Majorana}, and localised or quasilocalised modes in
braneworld constructions \cite{Linares2008WarpedBranes}. Scattering,
Green-function, and functional-determinant methods provide a general
framework for quantum vacuum effects in inhomogeneous backgrounds
\cite{GrahamEtAl2002VacuumEnergies,KennethKlich2008TGTG}.

Here we show that an isolated normalisable localised zero mode can
parametrically change the large-distance Casimir interaction in a
theory whose continuum spectrum of particles remains massive. It opens an infrared channel that
dominates over the direct exchange of massive quantum fluctuations at
large separation. We demonstrate this analytically for a non-negative
fluctuation operator whose only discrete state below the massive
continuum is the zero mode. The massive quantum fluctuations remain essential: at leading order their
local response dresses the bare coupling of each object to the zero-mode
fluctuations, while their direct exchange between the objects is
exponentially subleading.

We formulate the problem for semitransparent delta-function plates, but
the mechanism is not tied to this idealisation. It extends to general
localised repulsive objects with non-vanishing overlap with the zero
mode. For a planar background, the same transverse zero mode becomes a
gapless field propagating along the worldvolume, changing the
dimensional dependence of the interaction without altering the
underlying mechanism.

Topological defects provide a natural, but not necessary, realisation
of this mechanism. Once the position of a kink or domain wall is fixed,
the quadratic fluctuation operator is no longer translationally
invariant, while the translational symmetry of the underlying theory is
encoded in a normalisable zero mode
\cite{MantonSutcliffe2004}. We illustrate the effect with the
sine-Gordon kink, whose quadratic fluctuation operator is the exactly
solvable P\"oschl--Teller system. In this case, the exact Casimir
interaction approaches the predicted zero-mode asymptotics at large
separation, while the enhancement is spatially localised around the
kink and disappears as the objects are moved into the asymptotic
massive vacuum.

%%%%%%%%%%%%%%%%%%%%%%%%%%%%%%%%%%%%%%%%%
\section{Dressed collective zero-mode channel}
\label{sec:universal}
%%%%%%%%%%%%%%%%%%%%%%%%%%%%%%%%%%%%%%%%%

Consider the one-dimensional transverse fluctuation operator associated with a
static classical background,
\begin{equation}
K_{\rm bg}
=
-\frac{d^2}{dx^2}+m^2+V(x),
\qquad
\lim_{|x|\to\infty}V(x)=0,
\label{eq:qft-Kbg}
\end{equation}
with $m>0$ and $V(x)$ real. We take $K_{\rm bg}$ to be self-adjoint and non-negative, and
consider the spectral class
\begin{equation}
\sigma(K_{\rm bg})=\{0\}\cup[m^2,\infty),
\label{eq:qft-spectrum}
\end{equation}
where $\sigma(K_{\rm bg})$ denotes the spectrum of $K_{\rm bg}$. The zero
mode is normalisable and non-degenerate,
\begin{equation}
K_{\rm bg}\psi_0=0,
\qquad
\int_{-\infty}^{\infty}dx\,\psi_0^2(x)=1.
\label{eq:qft-zero-mode}
\end{equation}

For simplicity, and to make the mechanism explicit, we consider two repulsive
point-like objects described by Dirac delta potentials,
\begin{equation}
W(x)
=
\lambda_1\delta(x-a_1)+\lambda_2\delta(x-a_2),
\quad
\lambda_i>0,
\quad
a_{1,2}=c\mp\frac{d}{2}.
\label{eq:qft-Wdelta}
\end{equation}
Here $d>0$ is their separation and $c\in\mathbb{R}$ locates the centre of the
pair relative to the background. We regard the objects as fixed external
scatterers. The quantum fluctuations $\eta(\tau,x)$ propagate in their
presence, with the scattering encoded by the potential $W$ in the
quadratic fluctuation theory. The corresponding Euclidean action is
\begin{equation}
S^{(2)}[\eta]
=
\frac12\int d\tau\,dx\,
\left[
(\partial_\tau\eta)^2
+\eta K_{\rm bg}\eta
+W(x)\eta^2
\right].
\label{eq:qft-action}
\end{equation}
This quadratic action defines the model. It is not obtained by expanding a microscopic coupling to the full nonlinear field, which would generally generate a tadpole and classical backreaction; a microscopic realisation in which no tadpole arises is given in Sec.~\ref{subsec:PT-microscopic}. The corresponding linear field equation is\begin{equation}
\left[
-\partial_\tau^2+K_{\rm bg}+W(x)
\right]\eta(\tau,x)=0.
\label{eq:qft-eom}
\end{equation}
At fixed Euclidean frequency $\xi$, Fourier transformation in $\tau$,
together with Eq.~\eqref{eq:qft-Wdelta}, gives
\begin{equation}
(K_{\rm bg}+\xi^2)\eta_\xi(x)
+
\sum_{j=1}^{2}
\lambda_j\delta(x-a_j)\eta_\xi(a_j)
=0.
\label{eq:qft-eom-xi}
\end{equation}
This frequency-space equation will be our starting point: we shall separate
its zero-mode and massive components and take both into account.

The background Green function is defined by
\begin{equation}
(K_{\rm bg}+\xi^2)G_\xi(x,y)=\delta(x-y).
\label{eq:qft-Gdef}
\end{equation}
Its spectral decomposition separates the zero-mode pole exactly,
\begin{equation}
G_\xi(x,y)
=
\frac{\psi_0(x)\psi_0(y)}{\xi^2}
+
G_\xi^\perp(x,y),
\label{eq:qft-Gsplit}
\end{equation}
where
\begin{align}
(K_{\rm bg}+\xi^2)G_\xi^\perp(x,y)
&=
\delta(x-y)-\psi_0(x)\psi_0(y),
\label{eq:qft-Gperp-eq}
\\
\int dx\,\psi_0(x)G_\xi^\perp(x,y)
&=0.
\label{eq:qft-Gperp-orth}
\end{align}
The zero-mode term is the only infrared singularity;
$G_\xi^\perp$ is regular at $\xi=0$ because all remaining fluctuation
modes belong to the massive continuum starting at $m^2$.
Correspondingly, the fluctuation field can be decomposed as
\begin{equation}
\eta(\tau,x)
=
q(\tau)\psi_0(x)+\eta_\perp(\tau,x),
\quad
\int dx\,\psi_0(x)\eta_\perp(\tau,x)=0.
\label{eq:qft-field-split}
\end{equation}
Equation~\eqref{eq:qft-field-split} is an exact spectral decomposition
of the fluctuation field into the zero-mode and orthogonal sectors \cite{GoldstoneJackiw1975,GervaisSakita1975,Rajaraman1982,MantonSutcliffe2004}.
Substitution into Eq.~\eqref{eq:qft-action} gives
\begin{equation*}
\begin{aligned}
S^{(2)}[q,\eta_\perp]
={}&\frac12\int d\tau\,\biggl\{
\dot q^{\,2}\\
&+\int dx\,\bigl[(\partial_\tau\eta_\perp)^2
+\eta_\perp K_{\rm bg}\eta_\perp\bigr]\\
&+\sum_{j=1}^{2}\lambda_j
\bigl[q\psi_0(a_j)+\eta_\perp(\tau,a_j)\bigr]^2
\biggr\}.
\end{aligned}
\end{equation*}

The kinetic and $K_{\rm bg}$ cross terms vanish by orthogonality and
$K_{\rm bg}\psi_0=0$. The external potential $W$ mimicking the two objects, however, mixes
$q$ and $\eta_\perp$. This mixing will not be neglected.

We first present the large-separation infrared behaviour of the
interaction, which will be derived below from the full quadratic theory. 
For $d\gg m^{-1}$, direct propagation of massive fluctuations between
the two objects, represented by the delta-function terms in $W$, is
exponentially suppressed. The leading interaction is mediated by
fluctuations associated with the localised zero mode. Nevertheless,
the massive sector remains essential: through the mixing induced by
$W$, its local response dresses the coupling of each object to the
zero-mode fluctuations.
For each $i=1,2$, we consider the single-object configuration
obtained by retaining only the term $\lambda_i\delta(x-a_i)$ in
$W$ of Eq.~\eqref{eq:qft-Wdelta}. We denote by $\bar\alpha_i$
the resulting dressed infrared coupling of this object to the
zero-mode fluctuations. To leading infrared order, the
corresponding effective action for $q(\tau)$ takes the form
\begin{equation}
S_{\rm IR}^{(i)}[q]
=
\frac12\int d\tau\,
\left[
\dot q^{\,2}+\bar\alpha_i q^2
\right],
\label{eq:qft-SIR-single}
\end{equation}
whereas for the joint two-object configuration it becomes
\begin{equation}
S_{\rm IR}^{(12)}[q]
=
\frac12\int d\tau\,
\left[
\dot q^{\,2}+(\bar\alpha_1+\bar\alpha_2)q^2
\right].
\label{eq:qft-SIR-pair}
\end{equation}
The zero-mode amplitude therefore acquires the frequencies
\begin{equation}
\Omega_i=\sqrt{\bar\alpha_i},
\qquad
\Omega_{12}=\sqrt{\bar\alpha_1+\bar\alpha_2}.
\label{eq:qft-omegas}
\end{equation}
The infrared interaction energy is obtained as the relative zero-point
energy of the joint configuration with respect to the two isolated
one-object configurations. Since the unperturbed zero mode has zero
frequency,
\begin{equation}
E_{\rm IR}(c,d)
=
\frac12
\left[
\sqrt{\bar\alpha_1+\bar\alpha_2}
-\sqrt{\bar\alpha_1}
-\sqrt{\bar\alpha_2}
\right].
\label{eq:qft-Epreview}
\end{equation}
Equations~\eqref{eq:qft-SIR-single}--\eqref{eq:qft-Epreview}
anticipate the infrared behaviour of the interaction.
We now derive the exact interaction energy from the full quadratic
theory, retaining the mixing between the zero-mode and orthogonal
sectors, and establish Eq.~\eqref{eq:qft-Epreview} as its leading
large-distance contribution.

Writing the zero-mode amplitude in Fourier space as
\begin{equation}
q(\tau)
=
\int\frac{d\xi}{2\pi}\,e^{i\xi\tau}q_\xi,
\end{equation}
the field decomposition in Eq.~\eqref{eq:qft-field-split} becomes
\begin{equation}
\eta_\xi(x)
=
q_\xi\psi_0(x)+\eta_{\perp,\xi}(x).
\label{eq:qft-field-split-xi}
\end{equation}
Multiplying Eq.~\eqref{eq:qft-eom-xi} by $\psi_0(x)$ and integrating over
$x$ gives the zero-mode component,
\begin{equation}
\xi^2 q_\xi
+
\sum_{j=1}^{2}
\lambda_j\psi_0(a_j)\eta_\xi(a_j)
=
0.
\label{eq:qft-zero-eq}
\end{equation}
The orthogonal component $\eta_{\perp,\xi}$ contains the massive fluctuation
modes. Its equation can be solved using $G_\xi^\perp$, yielding
\begin{equation}
\eta_{\perp,\xi}(x)
=
-\sum_{j=1}^{2}
\lambda_j
G_\xi^\perp(x,a_j)\eta_\xi(a_j).
\label{eq:qft-perp-response}
\end{equation}
No fluctuation mode has been neglected in obtaining these equations.
Equation~\eqref{eq:qft-perp-response} gives the exact response of the massive
fluctuations to the field values at the two objects. Substituting this
response into Eq.~\eqref{eq:qft-zero-eq} accounts exactly for their effect on
the frequency-space dynamics of the zero-mode amplitude $q_\xi$.

Define
\begin{equation}
\psi_i\equiv\psi_0(a_i),
\qquad
g_{ij}^\perp(\xi)\equiv G_\xi^\perp(a_i,a_j).
\label{eq:qft-psi-g}
\end{equation}
For a single isolated object at $a_i$, evaluating
Eq.~\eqref{eq:qft-perp-response} at its position gives
\begin{equation}
\eta_\xi(a_i)
=
q_\xi\psi_i
-
\lambda_i g_{ii}^\perp(\xi)\eta_\xi(a_i),
\label{eq:qft-single-response}
\end{equation}
and therefore
\begin{equation}
\eta_\xi(a_i)
=
\frac{q_\xi\psi_i}
{1+\lambda_i g_{ii}^\perp(\xi)}.
\label{eq:qft-single-field}
\end{equation}
Point-supported interactions and their Green-function reduction are standard
\cite{Albeverio2005,MunozCastanedaEtAl2013DeltaPlates}.
Substituting Eq.~\eqref{eq:qft-single-field} into the zero-mode equation,
Eq.~\eqref{eq:qft-zero-eq}, shows that object $i$ enters through the
frequency-dependent dressed coupling
\begin{equation}
\alpha_i(\xi)
=
\frac{\lambda_i\psi_i^2}
{1+\lambda_i g_{ii}^\perp(\xi)},
\qquad
\bar\alpha_i\equiv\alpha_i(0).
\label{eq:qft-alpha}
\end{equation}
The numerator $\lambda_i\psi_i^2$ is the bare coupling to the zero mode,
whereas the denominator accounts for the local response of the massive
fluctuations at the position of the object. Thus $\alpha_i(\xi)$ is the
zero-mode coupling dressed by the massive fluctuations, and $\bar\alpha_i$
is its infrared limit.

With both objects present, Eq.~\eqref{eq:qft-perp-response} gives
\begin{align}
    [1+\lambda_1 g_{11}^\perp]\eta_\xi(a_1)
    +\lambda_2 g_{12}^\perp\eta_\xi(a_2)
    &=q_\xi\psi_1,
    \label{eq:two-object-field-values-1}
    \\
    \lambda_1 g_{21}^\perp\eta_\xi(a_1)
    +[1+\lambda_2 g_{22}^\perp]\eta_\xi(a_2)
    &=q_\xi\psi_2.
    \label{eq:two-object-field-values-2}
\end{align}
It is useful to collect this system in the matrix
\begin{equation}
    M_\perp(\xi)
    =
    \begin{pmatrix}
        1+\lambda_1g_{11}^\perp(\xi)
        &
        \lambda_2g_{12}^\perp(\xi)
        \\[2mm]
        \lambda_1g_{21}^\perp(\xi)
        &
        1+\lambda_2g_{22}^\perp(\xi)
    \end{pmatrix}.
    \label{eq:qft-Mperp}
\end{equation}
The exact frequency-dependent dressed coupling of the two-object
configuration to the zero-mode fluctuations is then
\begin{equation}
    \alpha_{12}(\xi)
    =
    \begin{pmatrix}
        \lambda_1\psi_1 & \lambda_2\psi_2
    \end{pmatrix}
    M_\perp^{-1}(\xi)
    \begin{pmatrix}
        \psi_1 \\ \psi_2
    \end{pmatrix}.
    \label{eq:qft-alpha12-exact}
\end{equation}
Indeed, solving Eqs.~\eqref{eq:two-object-field-values-1}
and~\eqref{eq:two-object-field-values-2} and substituting the result into
Eq.~\eqref{eq:qft-zero-eq} gives
\begin{equation}
    \left[\xi^2+\alpha_{12}(\xi)\right]q_\xi=0.
    \label{eq:qft-zero-eq-two-objects}
\end{equation}

The structure of $M_\perp$ separates the two roles played by the massive
fluctuations. Its diagonal entries contain the local propagation of virtual
massive fluctuations from an object back to the same object. These local
processes modify the bare coupling to the zero mode and determine the dressed
couplings $\alpha_i(\xi)$ in Eq.~\eqref{eq:qft-alpha}. By contrast, the
off-diagonal entries describe the propagation of virtual massive fluctuations
from one object to the other.

Under the decay assumptions specified in the Supplementary material, the mass
gap exponentially suppresses the propagation of virtual massive fluctuations
between well-separated objects \cite{CombesThomas1973}. After separating the
zero-mode pole, the reduced Green function satisfies, for any $0<\mu<m$,
\begin{equation}
    g_{12}^\perp(\xi)=O(e^{-\mu d}),
    \qquad 0<\mu<m,
    \label{eq:reduced-resolvent-offdiag}
\end{equation}
uniformly at low Euclidean frequencies. The corresponding estimate for the
present one-dimensional problem is derived in the Supplementary material.

Consequently,
Eqs.~\eqref{eq:two-object-field-values-1}
and~\eqref{eq:two-object-field-values-2}
become asymptotically local: to leading order, the field value at each object is given by the
corresponding isolated-object expression in Eq.~\eqref{eq:qft-single-field}.
Equation~\eqref{eq:qft-alpha12-exact} therefore gives, at low
Euclidean frequencies,
\begin{equation}
    \alpha_{12}(\xi)
    =
    \alpha_1(\xi)+\alpha_2(\xi)
    +O\!\left[
        \bigl(\alpha_1(\xi)+\alpha_2(\xi)\bigr)e^{-\mu d}
    \right].
    \label{eq:alpha12-asymptotic}
\end{equation}

Regularity of $G_\xi^\perp$ at $\xi=0$ and its smooth dependence on $\xi^2$
give the low-frequency expansion
\begin{equation}
    \alpha_i(\xi)
    =
    \bar\alpha_i[1+O(\xi^2)],
    \qquad \xi\to0.
    \label{eq:alpha-infrared}
\end{equation}
Thus the local response of the massive fluctuations remains at leading order
through $\bar\alpha_i$, whereas direct propagation of those fluctuations
between the two objects is exponentially subleading.

For point-like objects, the exact Gaussian determinant reduces to the field
values at their positions. Defining
\begin{equation}
    G_{ij}(\xi)
    \equiv
    G_\xi(a_i,a_j)
    =
    \frac{\psi_i\psi_j}{\xi^2}
    +
    g_{ij}^\perp(\xi),
    \label{eq:qft-Gij}
\end{equation}
the relative determinant of the two-object configuration with respect to the
two isolated objects is
\begin{equation}
    \Delta(\xi)
    =
    \frac{\det M(\xi)}
    {[1+\lambda_1G_{11}(\xi)]
     [1+\lambda_2G_{22}(\xi)]},
    \qquad
    M_{ij}(\xi)
    =
    \delta_{ij}+\lambda_jG_{ij}(\xi).
    \label{eq:qft-relative-determinant}
\end{equation}
This is the finite-rank form of the standard relative-determinant
construction for two scatterers
\cite{KennethKlich2008TGTG,MunozCaveroSancho2026PT}.

Using the decomposition in Eq.~\eqref{eq:qft-Gij}, the matrix determinant
lemma, and the couplings defined in Eqs.~\eqref{eq:qft-alpha}
and~\eqref{eq:qft-alpha12-exact}, the full relative determinant factorises
exactly as
\begin{equation}
    \Delta(\xi)
    =
    \Delta_\perp(\xi)
    \frac{
        \xi^2[\xi^2+\alpha_{12}(\xi)]
    }{
        [\xi^2+\alpha_1(\xi)]
        [\xi^2+\alpha_2(\xi)]
    },
    \label{eq:qft-full-factorisation}
\end{equation}
where the relative determinant of the massive sector is
\begin{equation}
    \Delta_\perp(\xi)
    =
    \frac{\det M_\perp(\xi)}
    {(M_\perp)_{11}(M_\perp)_{22}}
    =
    1-
    \frac{(M_\perp)_{12}(M_\perp)_{21}}
    {(M_\perp)_{11}(M_\perp)_{22}}.
    \label{eq:qft-Deltaperp}
\end{equation}

The second form in Eq.~\eqref{eq:qft-Deltaperp} makes the physical content
transparent. The off-diagonal entries of $M_\perp$ describe propagation of
virtual massive fluctuations between the two objects. A contribution mediated
entirely by the massive fluctuations therefore requires propagation from one
object to the other and back, and contains the product
$g_{12}^\perp g_{21}^\perp$. Hence
Eq.~\eqref{eq:reduced-resolvent-offdiag} gives
\begin{equation}
    \log\Delta_\perp(\xi)
    =
    O(e^{-2\mu d}).
    \label{eq:qft-Deltaperp-asymptotic}
\end{equation}

The exact one-loop interaction energy is
\begin{equation}
    E_{\rm int}
    =
    \frac{1}{2\pi}
    \int_0^\infty d\xi\,\log\Delta(\xi).
    \label{eq:qft-Eexact}
\end{equation}

The interaction energy is defined by
$E_{\rm int}=E_{12}-E_1-E_2+E_0$, with all four vacuum energies
evaluated in the same prescribed background and using a common
regularisation. This subtraction cancels the ultraviolet divergences
at one loop; the infrared convergence established below then ensures
that $E_{\rm int}$ is finite and well defined.

For repulsive objects, the couplings entering
Eq.~\eqref{eq:qft-full-factorisation} remain finite and positive at
$\xi=0$, so that $\Delta(\xi)\propto\xi^2$ and
$\log\Delta(\xi)\sim2\log\xi$. The infrared singularity is therefore
integrable. At large $\xi$, the relative determinant contains only
propagation between separated objects, which is exponentially suppressed,
ensuring ultraviolet convergence.

Under the same decay assumptions, at large separations, $d\gg m^{-1}$, with
the centre of the pair kept fixed relative to the background, the two objects
lie in the asymptotic tails of the localised zero mode. Since
\begin{equation}
    \psi_i^2=\psi_0^2(a_i)=O(e^{-md}),
\end{equation}
while the local massive-fluctuation response in the denominator of
Eq.~\eqref{eq:qft-alpha} remains finite, the infrared dressed couplings obey
\begin{equation}
    \bar\alpha_i=O(e^{-md}).
    \label{eq:qft-alpha-large-d}
\end{equation}
The characteristic Euclidean frequencies associated with these couplings are
therefore
\begin{equation}
    \xi
    \sim
    \sqrt{\bar\alpha_1+\bar\alpha_2}
    =
    O(e^{-md/2}).
    \label{eq:qft-xi-scale}
\end{equation}

On this infrared scale,
Eqs.~\eqref{eq:alpha12-asymptotic},
\eqref{eq:alpha-infrared}, and
\eqref{eq:qft-Deltaperp-asymptotic} reduce the full relative determinant
in Eq.~\eqref{eq:qft-full-factorisation} to
\begin{equation}
    \Delta_{\rm IR}(\xi)
    =
    \frac{
        \xi^2(\xi^2+\bar\alpha_1+\bar\alpha_2)
    }{
        (\xi^2+\bar\alpha_1)
        (\xi^2+\bar\alpha_2)
    }.
    \label{eq:qft-DeltaIR}
\end{equation}
The corrections to this infrared form give only subleading contributions to
the large-distance interaction energy; the corresponding estimate is given in
the Supplementary material.

The leading contribution can then be evaluated directly. Using
\begin{equation}
    \int_0^\infty d\xi\,
    \log\frac{\xi^2+A^2}{\xi^2+B^2}
    =
    \pi(A-B),
    \qquad A,B\geq0,
    \label{eq:qft-log-integral}
\end{equation}
in Eq.~\eqref{eq:qft-Eexact} gives
\begin{equation}
    E_{\rm int}(c,d)
    =
    \frac12
    \left[
        \sqrt{\bar\alpha_1+\bar\alpha_2}
        -\sqrt{\bar\alpha_1}
        -\sqrt{\bar\alpha_2}
    \right]
    +o(e^{-md/2}).
    \label{eq:qft-E-large-d}
\end{equation}

We have thus derived Eq.~\eqref{eq:qft-Epreview} as the leading
infrared contribution to the exact interaction energy at large
separations. This contribution is mediated by quantum fluctuations
associated with the localised zero mode. Virtual massive fluctuations
remain essential because their local response dresses the bare
zero-mode couplings into $\bar\alpha_i$, while their direct propagation
between the two objects is exponentially subleading
(see Supplementary material).

For a planar background in $(3+1)$-dimensional spacetime, the transverse
zero mode becomes a field $q(\tau,\mathbf y)$ propagating along the
two-dimensional spatial worldvolume. In the absence of the semitransparent
plates, it therefore defines a gapless worldvolume branch. In the presence
of one or both plates, the infrared effective actions are
\begin{equation}
S_{\rm IR}^{(i)}[q]
=
\frac12\int d\tau\,d^2\mathbf y\,
\left[
(\partial_\tau q)^2
+
|\boldsymbol\nabla_\parallel q|^2
+
\bar\alpha_i q^2
\right],
\label{eq:qft-planar-single}
\end{equation}
and
\begin{equation}
S_{\rm IR}^{(12)}[q]
=
\frac12\int d\tau\,d^2\mathbf y\,
\left[
(\partial_\tau q)^2
+
|\boldsymbol\nabla_\parallel q|^2
+
(\bar\alpha_1+\bar\alpha_2)q^2
\right].
\label{eq:qft-planar-pair}
\end{equation}
The corresponding worldvolume dispersions are
\begin{equation}
\omega_i^2
=
\mathbf k_\parallel^2+\bar\alpha_i,
\qquad
\omega_{12}^2
=
\mathbf k_\parallel^2+\bar\alpha_1+\bar\alpha_2.
\label{eq:qft-planar-dispersions}
\end{equation}

After Fourier transformation in Euclidean time and in the directions
parallel to the plates, the worldvolume momentum enters through its
Euclidean magnitude
\begin{equation}
p\equiv\sqrt{\xi^2+\mathbf k_\parallel^2}.
\label{eq:qft-p-def}
\end{equation}
The transverse problem is therefore the same as in $(1+1)$ dimensions with
$\xi$ replaced by $p$, and its large-distance infrared determinant is
\begin{equation}
\Delta_{\rm IR}(p)
=
\frac{
p^2(p^2+\bar\alpha_1+\bar\alpha_2)
}{
(p^2+\bar\alpha_1)(p^2+\bar\alpha_2)
}.
\label{eq:qft-planar-DeltaIR}
\end{equation}
The leading interaction energy per unit plate area $S$ is consequently
\begin{equation}
\frac{E_{\rm int}}{S}
=
\frac{1}{4\pi^2}
\int_0^\infty dp\,p^2
\log\Delta_{\rm IR}(p).
\label{eq:qft-planar-energy-integral}
\end{equation}
This integral can be evaluated directly. For $a,b>0$,
\begin{equation*}
\int_0^\infty dp\,p^2
\log
\frac{p^2(p^2+a+b)}
{(p^2+a)(p^2+b)}
=
\frac{\pi}{3}
\left[
a^{3/2}+b^{3/2}-(a+b)^{3/2}
\right].
\end{equation*}
Hence
\begin{equation}
\frac{E_{\rm int}}{S}
=
\frac{1}{12\pi}
\left[
\bar\alpha_1^{3/2}
+
\bar\alpha_2^{3/2}
-
(\bar\alpha_1+\bar\alpha_2)^{3/2}
\right]
+
o(e^{-3md/2}).
\label{eq:qft-planar-energy}
\end{equation}
Since $\bar\alpha_i=O(e^{-md})$, the leading planar interaction therefore
decays as $e^{-3md/2}$.

The physical mechanism is unchanged. The interaction is governed at large
separation by the gapless worldvolume fluctuations generated by the
localised transverse zero mode, while the local response of the virtual massive fluctuations dresses the
bare zero-mode couplings into $\bar\alpha_i$.
Direct propagation of massive fluctuations between the two plates remains
exponentially subleading after the worldvolume momentum integration
(see Supplementary material).

Finally, the mechanism is not specific to point-like objects. For a
non-negative localised object described by a potential $W_i(x)$ of compact
support, the frequency-dependent dressed coupling to the zero-mode
fluctuations generalises to
\begin{equation}
    \alpha_i(\xi)
    =
    \int dx\,dy\,
    \psi_0(x)W_i(x)
    \left[1+G_\xi^\perp W_i\right]^{-1}(x,y)
    \psi_0(y).
    \label{eq:qft-alpha-extended}
\end{equation}
For $W_i(x)=\lambda_i\delta(x-a_i)$,
Eq.~\eqref{eq:qft-alpha-extended} reduces directly to
Eq.~\eqref{eq:qft-alpha}.

For two localised objects of fixed finite extent and large separation, the
same exponential decay of $G_\xi^\perp(x,y)$ suppresses direct propagation
of virtual massive fluctuations between their supports. Their local
massive-fluctuation response remains encoded in the dressed couplings
$\alpha_i(\xi)$. The long-distance mechanism derived above therefore does
not rely on the point-like approximation.

%%%%%%%%%%%%%%%%%%%%%%%%%%%%%%%%%%%%%%%%%
\section{Topological realisation: the sine-Gordon kink}
\label{sec:PT-realisation}
%%%%%%%%%%%%%%%%%%%%%%%%%%%%%%%%%%%%%%%%%

Topological kinks provide a natural realisation of the spectral setting
introduced in Sec.~\ref{sec:universal}. Consider a real scalar field with
action
\begin{equation}
    S[\phi]
    =
    \int \dd t\,\dd x\,
    \left[
        \frac12(\partial_t\phi)^2
        -
        \frac12(\partial_x\phi)^2
        -
        U(\phi)
    \right].
    \label{eq:scalar-action}
\end{equation}
A static kink \(\phi_K(x)\) satisfies
\[
    -\phi_K''+U'(\phi_K)=0,
\]
and its quadratic fluctuation operator is
\begin{equation}
    \Kk
    =
    -\frac{d^2}{dx^2}
    +
    U''(\phi_K).
    \label{eq:KK}
\end{equation}
Differentiating the static field equation gives
\[
    \Kk\phi_K'=0.
\]
The translational zero mode is therefore
\begin{equation}
    \psi_0(x)
    =
    {\cal N}\phi_K'(x),
    \label{eq:psi0}
\end{equation}
with \({\cal N}\) fixed by the normalisation in
Eq.~\eqref{eq:qft-zero-mode}. If the two asymptotic vacua have the same
curvature, \(U''(v_\pm)=m^2>0\), this zero mode lies below a massive
continuum. A generic kink may also possess additional bound states, so the
spectral class of Sec.~\ref{sec:universal} applies when the translational
zero mode is the only discrete state below the continuum
\cite{Rajaraman1982,Coleman1985,MantonSutcliffe2004}.

In this realisation, the classical kink provides the prescribed
background for the quantum fluctuations. These fluctuations scatter
from the two fixed external probes, represented by $W$ in the
quadratic action~\eqref{eq:qft-action}, and thereby mediate their
Casimir interaction. The translational zero mode contributes to
this interaction through quantum fluctuations of its amplitude $q$.
Our calculation is restricted to the quadratic fluctuation theory
around the classical kink, rather than the full nonlinear
sine-Gordon theory.

For the sine-Gordon potential
\[
    U_{\rm sG}(\phi)=1-\cos\phi,
\]
the kink is
\[
    \phi_K(x)=4\arctan e^x.
\]
In units where the asymptotic fluctuation mass is \(m=1\), the quadratic
fluctuation operator is the one-soliton P\"oschl--Teller operator
\begin{equation}
    \Kpt
    =
    -\frac{d^2}{dx^2}
    +
    1-2\sech^2x,
    \label{eq:KPT}
\end{equation}
with normalised zero mode
\begin{equation}
    \psi_0(x)
    =
    \frac{1}{\sqrt2}\sech x,
    \qquad
    \Kpt\psi_0=0.
    \label{eq:PT-zero-mode}
\end{equation}
Its spectrum is
\begin{equation}
    \sigma(\Kpt)
    =
    \{0\}\cup[1,\infty),
    \label{eq:PT-spectrum}
\end{equation}
so this background belongs exactly to the spectral class considered in
Sec.~\ref{sec:universal}.

For the P\"oschl--Teller background, the exact Green function is known
\cite{MunozCaveroSancho2026PT}. After subtraction of the zero-mode pole,
its diagonal reduced resolvent at zero frequency is
\begin{equation}
    g_{aa}^\perp(0)
    =
    \frac12-\frac14\sech^2 a .
    \label{eq:PT-gperp}
\end{equation}
Using Eq.~\eqref{eq:qft-alpha} together with
Eq.~\eqref{eq:PT-zero-mode}, the dressed infrared coupling of a
delta-function plate of strength \(\lambda\) at \(a\) is therefore
\begin{equation}
    \bar\alpha(a)
    =
    \frac{
        (\lambda/2)\sech^2 a
    }{
        1+\lambda
        \left(
            \frac12-\frac14\sech^2 a
        \right)
    }.
    \label{eq:PT-alpha-single}
\end{equation}
For two identical plates at $a_{1,2}=c\mp d/2$, the corresponding dressed
infrared couplings are simply $\bar\alpha_i(c,d)=\bar\alpha(a_i)$.
Thus the sine-Gordon kink provides an exactly solvable realisation in
which the spatial dependence of the dressed infrared couplings is fixed
explicitly by the localised zero-mode profile and by the local response of
the massive fluctuations. The exact interaction energy, by
contrast, is obtained from Eq.~\eqref{eq:qft-Eexact} using the full
P\"oschl--Teller Green function.

\subsection{Microscopic sigma-model realisation}
\label{subsec:PT-microscopic}

The preceding sine-Gordon example treats the impurities within
a prescribed quadratic theory. A microscopic coupling $W(x)F(\phi)$ to the nonlinear scalar field would generally produce a tadpole when $F'(\phi_K(a_i))\neq 0$ at the impurity positions, thereby modifying the classical background. To show that this is not inevitable,
consider an $\mathbb S^2$ nonlinear sigma model with
constrained field $\bm N=(N_x,N_y,N_z)$, $\bm N^2=4$.
Related sigma-model kinks were studied in
Ref.~\cite{PRL2008_131602}, with their stability and
fluctuation spectra analysed further in
Ref.~\cite{AlonsoIzquierdoEtAl2009Sigma}.
We take
\begin{equation}
\begin{aligned}
S[\bm N]=\frac1{2g^2}\int\dd t\,\dd x\,
\bigl[&\partial_\mu\bm N\cdot\partial^\mu\bm N\\
&-(4-N_z^2)-W(x)N_y^2\bigr],
\end{aligned}
\label{eq:PT-sigma-action}
\end{equation}
where $W$ is given by Eq.~\eqref{eq:qft-Wdelta}.
Setting $N_y=0$ and
$(N_x,N_z)=2(\sin(\phi/2),\cos(\phi/2))$
recovers sine-Gordon. Its kink becomes
$\bm N_K=(2\sech x,0,-2\tanh x)$.
Since the impurity energy and its first variation vanish
on this wall, it remains an exact classical solution
without tadpoles or backreaction. For repulsive impurities,
the wall also saturates the same classical energy bound
with zero, one or two impurities, giving
$E_{\rm int}^{\rm cl}=0$
(see~\ref{app:sigma-realisation}).

To identify the fluctuation operator, introduce the regular
transverse field $N_y=g\eta$ and the longitudinal fluctuation
$\phi=\phi_K+g\zeta$. The quadratic action contains no
$\zeta$--$\eta$ mixing, and its transverse Hessian is
\begin{equation}
\begin{aligned}
K_\eta
&=-\partial_x^2+\cos^2\frac{\phi_K}{2}
-\frac14(\phi_K')^2+W(x)\\
&=-\partial_x^2+1-2\sech^2x
+\sum_{i=1}^2\lambda_i\delta(x-a_i)\\
&=\Kpt+W(x).
\end{aligned}
\label{eq:PT-sigma-Hessian}
\end{equation}
Here we used $\phi_K'=2\sech x$ and
$\cos(\phi_K/2)=-\tanh x$.
The longitudinal block is independent of the impurities
and cancels in the interaction subtraction when the
wall centre is held fixed. Consequently, the complete
one-loop relative determinant is precisely
Eq.~\eqref{eq:qft-relative-determinant}, with interaction
energy~\eqref{eq:qft-Eexact} and dressed coupling
\eqref{eq:PT-alpha-single}. Hence the one-loop results of
Sec.~\ref{sec:enhancement} apply verbatim to this model.

In this microscopic realisation, the localised zero mode
describes an internal rotation of the wall, rather than
its translation. The model also admits an effective
interpretation as an easy-axis antiferromagnetic domain
wall, with $\bm n=\bm N/2$ the N\'eel order parameter
and $W(x)N_y^2/2$ representing local transverse
hard-axis anisotropy \cite{Haldane,Tveten}.
The angular zero mode requires axial symmetry of the
impurity-free theory; a uniform hard-axis anisotropy
would lift it. The correspondence is exact at Gaussian
order, not for generic impurities coupled to a single
sine-Gordon field or for the full nonlinear large-distance
limit at fixed $g$.

%%%%%%%%%%%%%%%%%%%%%%%%%%%%%%%%%%%%%%%%%
\section{Zero-mode enhancement of massive Casimir forces}
\label{sec:enhancement}
%%%%%%%%%%%%%%%%%%%%%%%%%%%%%%%%%%%%%%%%%

We now compare the exact interaction in the P\"oschl--Teller background,
its infrared asymptotics derived in Sec.~\ref{sec:universal}, and the
interaction between the same semitransparent plates in the homogeneous
massive vacuum. We use the dimensionless variables of
Sec.~\ref{sec:PT-realisation}, so that \(m=1\), and write
\[
    \kappa_\xi=\sqrt{1+\xi^2}.
\]

For two identical delta-function plates of strength \(\lambda>0\)
separated by \(d\), the homogeneous massive interaction is
\cite{MunozCastanedaEtAl2013DeltaPlates}
\begin{equation}
    E_{\rm m}^{(0)}(d)
    =
    \frac{1}{2\pi}
    \int_0^\infty \dd\xi\,
    \log\!\left[
        1-
        \left(
            \frac{\lambda}{2\kappa_\xi+\lambda}
        \right)^2
        e^{-2d\kappa_\xi}
    \right].
    \label{eq:massive-benchmark}
\end{equation}
The corresponding force,
\(F_{d,{\rm m}}^{(0)}=-\partial_dE_{\rm m}^{(0)}\), behaves at large
separation as
\begin{equation}
    F_{d,{\rm m}}^{(0)}(d)
    =
    O\!\left(d^{-1/2}e^{-2d}\right),
    \qquad
    d\to\infty.
    \label{eq:massive-asymptotics}
\end{equation}

For plates centred on the kink, \(c=0\), the two dressed infrared
couplings coincide. Equation~\eqref{eq:PT-alpha-single} gives
\begin{equation}
    \bar\alpha_1(0,d)
    =
    \bar\alpha_2(0,d)
    \equiv
    \bar\alpha(d)
    =
    \frac{
        (\lambda/2)\sech^2(d/2)
    }{
        1+\lambda
        \left[
            \frac12-\frac14\sech^2(d/2)
        \right]
    }.
    \label{eq:centered-alpha}
\end{equation}
Equation~\eqref{eq:qft-Epreview} then gives the infrared interaction
\begin{equation}
    E_{\rm IR}(0,d)
    =
    \left(
        \frac{1}{\sqrt2}-1
    \right)
    \sqrt{\bar\alpha(d)},
    \label{eq:centered-Ezm}
\end{equation}
with force
\begin{equation}
    F_{d,{\rm IR}}(0,d)
    =
    -\partial_d E_{\rm IR}(0,d).
    \label{eq:centered-Fzm}
\end{equation}
Since
\[
    \bar\alpha(d)
    =
    \frac{4\lambda}{\lambda+2}\,
    e^{-d}
    [1+O(e^{-d})],
\]
we obtain
\begin{equation}
    F_{d,{\rm IR}}(0,d)
    \sim
    -
    \left(
        1-\frac{1}{\sqrt2}
    \right)
    \sqrt{\frac{\lambda}{\lambda+2}}\,
    e^{-d/2}.
    \label{eq:centered-Fzm-asymptotics}
\end{equation}

For two identical plates in the weak-coupling regime,
Eq.~\eqref{eq:centered-Fzm-asymptotics} gives
$F_{d,{\rm IR}}\propto\lambda^{1/2}$, whereas the homogeneous
massive force obtained from Eq.~\eqref{eq:massive-benchmark}
starts at $O(\lambda^2)$. This non-analytic behaviour arises
from the probe-induced lifting of the zero mode and cannot
be captured by an ordinary weak-coupling perturbative
expansion. In the microscopic realisation of
Sec.~\ref{subsec:PT-microscopic} this additionally requires
$\lambda\gg g^{4}e^{d}$, up to $O(1)$ factors, so that the
angular zero mode remains localised
(see~\ref{app:sigma-realisation}).

Thus the infrared force associated with the zero-mode channel decays
as $e^{-d/2}$, parametrically more slowly than the homogeneous
massive benchmark $d^{-1/2}e^{-2d}$.

\begin{figure}[!t]
    \centering
    \includegraphics[width=\columnwidth]
    {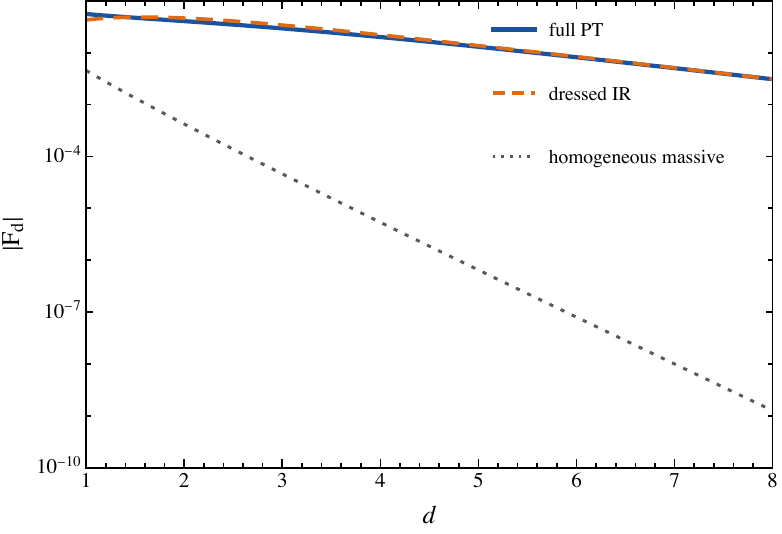}
    \caption{Absolute value of the interplate force for two identical repulsive
plates with $\lambda=1$ centred on the kink. The dressed infrared
approximation approaches the exact one-loop P\"oschl--Teller force at large
separation, whereas the homogeneous massive force decays parametrically
faster.}
    \label{fig:PT-force-comparison}
\end{figure}

To verify this explicitly, let \(E_{\rm PT}(c,d)\) denote the exact
energy obtained from Eq.~\eqref{eq:qft-Eexact} using the
P\"oschl--Teller Green function of
Ref.~\cite{MunozCaveroSancho2026PT}, and define
\[
    F_{d,{\rm PT}}(c,d)
    =
    -\partial_d E_{\rm PT}(c,d).
\]
Figure~\ref{fig:PT-force-comparison} compares this exact force for
\(\lambda=1\) and \(c=0\) with the infrared prediction
\eqref{eq:centered-Fzm} and the homogeneous massive benchmark.

The agreement of the exact P\"oschl--Teller force with the infrared
prediction confirms the asymptotic reduction derived from the full
Gaussian determinant. The massive sector has not been discarded: its
local response is contained in \(\bar\alpha_i\), while direct propagation
through that sector is subleading at large separation.

To quantify the enhancement of the complete interaction, we define
\begin{equation}
    A(c,d)
    =
    \frac{
        |F_{d,{\rm PT}}(c,d)|
    }{
        |F_{d,{\rm m}}^{(0)}(d)|
    }.
    \label{eq:Aamp}
\end{equation}

\begin{figure}[!t]
    \centering
    \includegraphics[width=\columnwidth]
    {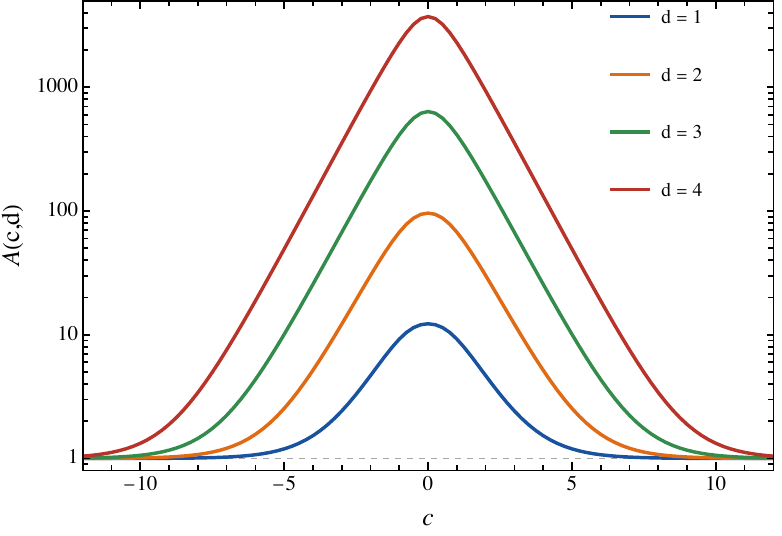}
    \caption{Spatial localisation of the exact amplification factor
    \(A(c,d)\) for \(\lambda=1\) and several fixed separations. The
    enhancement is maximal when the pair overlaps the kink region and
    approaches the homogeneous baseline \(A=1\) as both plates are
    displaced into the asymptotic vacuum.}
    \label{fig:PT-localised-enhancement}
\end{figure}

Figure~\ref{fig:PT-localised-enhancement} shows that the enhancement is
localised around the kink. At fixed \(d\), it is largest where the plates
overlap appreciably with the zero-mode profile. As \(|c|\to\infty\), both
plates probe the asymptotic homogeneous vacuum and \(A(c,d)\to1\).
Thus \(c\) resolves the spatial localisation of the additional infrared
channel, while \(d\) exposes its slower large-distance decay.

This comparison concerns an asymptotically massive homogeneous theory.
A genuinely massless homogeneous field instead possesses an extended
gapless sector and exhibits the usual algebraic large-distance Casimir
behaviour.

%%%%%%%%%%%%%%%%%%%%%%%%%%%%%%%%%%%%%%%%%
\section{Conclusions}
\label{sec:conclusions}
%%%%%%%%%%%%%%%%%%%%%%%%%%%%%%%%%%%%%%%%%

We have shown that a normalisable localised zero mode opens an
additional Casimir interaction channel whose decay is parametrically
slower than that associated with the exchange of massive quantum
fluctuations, while the fluctuation field still possesses a massive
continuous sector. The massive fluctuations remain essential: their
local response dresses the coupling of each object to the zero-mode
fluctuations, while their direct exchange between distant objects is
exponentially subleading.

For backgrounds satisfying the spectral condition
$\sigma(K_{\rm bg})=\{0\}\cup[m^2,\infty)$ and the short-range
assumption~\eqref{app:eq:short-range}, the leading large-distance
interaction at fixed $c$ in $1+1$ dimensions is
\[
	E_{\rm int}(c,d)
	=
	\frac12
	\left[
		\sqrt{\bar\alpha_1+\bar\alpha_2}
		-
		\sqrt{\bar\alpha_1}
		-
		\sqrt{\bar\alpha_2}
	\right]
	+
	o\!\left(e^{-md/2}\right),
\]
where \(\bar\alpha_i\) are the dressed infrared couplings of the two
objects. At fixed \(c\), localisation of the zero mode gives
\(\bar\alpha_i=O(e^{-md})\), and therefore
\(E_{\rm int}=O(e^{-md/2})\). In a planar \((3+1)\)-dimensional
setting, the same transverse zero mode generates a gapless worldvolume
branch and the leading interaction instead scales as
\(e^{-3md/2}\).

The sine-Gordon kink provides a topological background whose
exactly solvable fluctuation operator realises this mechanism
through its translational zero mode. For
two identical plates centred on the kink, the leading infrared force
decays as \(e^{-d/2}\), whereas the corresponding homogeneous massive
force behaves as \(d^{-1/2}e^{-2d}\). The exact Casimir force in the
P\"oschl--Teller background approaches this asymptotic behaviour at
large separation, demonstrating explicitly the dominance of the
zero-mode interaction channel. The enhancement is spatially localised
around the kink and tends to the homogeneous baseline \(A=1\) as the
pair is moved into the asymptotic vacuum.

A microscopic realisation of the same Gaussian problem, based on an
$\mathbb S^2$ nonlinear sigma model in which the impurities couple to
the transverse field and no tadpole or classical backreaction arises,
is given in Sec.~\ref{subsec:PT-microscopic}. There the localised zero
mode is an internal rotation of the wall rather than its translation,
and the one-loop law holds in a joint semiclassical, large-distance
regime, $g^{2}e^{md/2}\ll1$ up to coupling-dependent factors, where
$g$ is the coupling of Eq.~\eqref{eq:PT-sigma-action}
(see~\ref{app:sigma-realisation}); it is not asserted as an exact
asymptotic of the nonlinear theory at fixed $g$.

The principal conclusion is that the presence of a massive continuous
sector does not by itself determine the infrared behaviour of the
Casimir interaction. A normalisable localised zero mode can open an
additional Casimir interaction channel that becomes parametrically
dominant at large separation over the direct exchange of massive
quantum fluctuations, although the interaction remains exponentially
suppressed. The mechanism does not require topology. Topological defects provide a
natural realisation through their zero modes: translational for the
sine-Gordon kink and, in the realisation of
Sec.~\ref{subsec:PT-microscopic}, an internal rotation.

\section*{Data availability}

No external datasets were used for the research described in this
article. The numerical results shown in the figures were generated
directly from the analytical expressions evaluated in this work.

\section*{Declaration of generative AI and AI-assisted technologies
in the manuscript preparation process}

During the preparation of this work, the authors used ChatGPT
(OpenAI) and Claude (Anthropic) to assist with language editing and \LaTeX{} coding. After
using this tool, the authors reviewed and edited the content as needed
and take full responsibility for the content of the article.

\section*{Acknowledgements}

J.M.C. and G.S.G. acknowledge partial financial support from Grant PID2021-123251NB-I00 funded by MCIN/AEI/10.13039/501100011033 and by the European Union, and from the Department of Education of the Junta de Castilla y León and FEDER Funds (Reference No.~CLU-2025-1-02-IMUVA). G.S.G. also acknowledges financial support from a University of Valladolid predoctoral contract co-funded by Banco Santander (Grant No.~CONTPR-2024-310) and from the Formación de Profesorado Universitario (FPU) programme of the Spanish Ministry of Science, Innovation and Universities (Grant No.~FPU24/01715). I.C.P. acknowledges partial financial support from Spanish MINECO/FEDER Grant PID2024-160228NB-I00, funded by MCIN/AEI/10.13039/501100011033 and the European Regional Development Fund (ERDF, ``A way of making Europe''), and by Grant E21-23R funded by the Government of Aragon and the European Union.

The authors thank Juan Mateos Guilarte, Jesús Clemente Gallardo, Manuel Donaire, and Beatriz Gómez López for valuable discussions.

%%%%%%%%%%%%%%%%%%%%%%%%%%%%%%%%%%%%%%%%%%%%%%%%%%%%
\appendix
\section{Supplementary material}
%%%%%%%%%%%%%%%%%%%%%%%%%%%%%%%%%%%%%%%%%%%%%%%%%%%%

\subsection{Reduced resolvent: infrared regularity and spatial decay}
\label{app:reduced-resolvent}

We collect here the properties of the reduced Green function used in
Sec.~\ref{sec:universal}. In addition to the spectral assumptions of
the main text, we assume that the real background potential is
exponentially short-ranged,
\begin{equation}
    |V(x)|\leq C e^{-\beta |x|},
    \qquad
    \beta>0.
    \label{app:eq:short-range}
\end{equation}
We consider \(a_{1,2}=c\mp d/2\), with \(c\) fixed and
\(d\to\infty\).

Let
\(\kappa_\xi=\sqrt{m^2+\xi^2}\), and let
\(f_\pm(x,\kappa)\) be the Jost solutions of
\[
    \left[
        -\frac{d^2}{dx^2}
        +V(x)+\kappa^2
    \right]f=0,
\]
normalised by
\(f_-(x,\kappa)\sim e^{\kappa x}\) as \(x\to-\infty\) and
\(f_+(x,\kappa)\sim e^{-\kappa x}\) as \(x\to+\infty\).
For \(x<y\),
\begin{equation}
    G_\xi(x,y)
    =
    \frac{
        f_-(x,\kappa_\xi)f_+(y,\kappa_\xi)
    }{
        {\cal W}(\kappa_\xi)
    },
    \label{app:eq:jost-green}
\end{equation}
where ${\cal W}(\kappa)=f_-'(x,\kappa)f_+(x,\kappa)
-f_-(x,\kappa)f_+'(x,\kappa)$ is their
$x$-independent Wronskian.

The normalisable zero mode corresponds to a simple zero of
\({\cal W}(\kappa)\) at \(\kappa=m\). Since
\[
    \kappa_\xi-m
    =
    \frac{\xi^2}{2m}
    +O(\xi^4),
\]
the Laurent expansion of Eq.~\eqref{app:eq:jost-green} separates as
\begin{equation}
    G_\xi(x,y)
    =
    \frac{\psi_0(x)\psi_0(y)}{\xi^2}
    +
    G_\xi^\perp(x,y),
    \label{app:eq:pole-subtraction}
\end{equation}
with \(G_\xi^\perp\) regular in \(\xi^2\) at the origin. Hence
\[
    g_{ii}^\perp(\xi)
    =
    g_{ii}^\perp(0)+O(\xi^2).
\]
Since \(G_\xi^\perp\) is non-negative on the orthogonal subspace and
\(\lambda_i>0\), the denominator in Eq.~\eqref{eq:qft-alpha} stays
bounded away from zero, and therefore
\begin{equation}
    \alpha_i(\xi)
    =
    \bar\alpha_i
    \left[1+O(\xi^2)\right],
    \qquad
    \xi\to0.
    \label{app:eq:alpha-regularity}
\end{equation}

After subtraction of the zero-mode pole, the Jost expansion is
regular at the threshold \(\kappa=m\). Evaluated at
\(a_{1,2}=c\mp d/2\), derivatives with respect to \(\kappa\) may
generate polynomial factors in \(d\), but the exponential decay remains
controlled by \(e^{-md}\). Hence, for every \(0<\mu<m\),
\begin{equation}
    G_\xi^\perp(a_1,a_2)
    =
    O(e^{-\mu d}),
    \qquad
    d\to\infty,
    \label{app:eq:reduced-kernel-bound}
\end{equation}
uniformly for \(\xi\) in a fixed neighbourhood of the origin.
Away from the threshold the ordinary Jost decay
\(e^{-\kappa_\xi d}\) is stronger.

It is essential that
Eq.~\eqref{app:eq:reduced-kernel-bound} refers to the
\emph{reduced} resolvent. The full Green function contains the
zero-mode pole
\(\psi_0(a_1)\psi_0(a_2)/\xi^2\) and therefore does not satisfy such
a bound uniformly as \(\xi\to0\).

\subsection{Large-distance reduction of the full determinant}
\label{app:infrared-reduction}

Equation~\eqref{app:eq:reduced-kernel-bound} makes the off-diagonal
entries of \(M_\perp\) \(O(e^{-\mu d})\), while its diagonal entries
remain bounded away from zero for repulsive objects. Direct inversion
of the \(2\times2\) matrix therefore gives
\begin{equation}
    \alpha_{12}(\xi)
    =
    \alpha_1(\xi)+\alpha_2(\xi)
    +
    O\!\left[
        \bigl(\alpha_1(\xi)+\alpha_2(\xi)\bigr)e^{-\mu d}
    \right],
    \label{app:eq:alpha12-bound}
\end{equation}
uniformly in the infrared regime. Likewise,
Eq.~\eqref{eq:qft-Deltaperp} gives
\begin{equation}
    \log\Delta_\perp(\xi)
    =
    O(e^{-2\mu d}).
    \label{app:eq:Deltaperp-bound}
\end{equation}

To control the integrated energy, set
\[
    s_d=\bar\alpha_1+\bar\alpha_2,
    \qquad
    r_i=\frac{\bar\alpha_i}{s_d},
    \qquad
    r_1+r_2=1.
\]
For fixed \(c\) and non-vanishing asymptotic zero-mode tails,
\(s_d=O(e^{-md})\), so the relevant frequency scale is
\(\xi=O(\sqrt{s_d})\).

With \(\xi=\sqrt{s_d}\,y\), Eqs.~\eqref{app:eq:alpha-regularity},
\eqref{app:eq:alpha12-bound}, and
\eqref{app:eq:Deltaperp-bound}, together with the exact factorisation
\eqref{eq:qft-full-factorisation}, give
\begin{equation}
    \Delta(\sqrt{s_d}\,y)
    =
    \frac{y^2(y^2+1)}
    {(y^2+r_1)(y^2+r_2)}
    [1+o(1)]
    \label{app:eq:scaled-determinant}
\end{equation}
for fixed \(y\).

This limit may be taken through the energy integral. Near \(y=0\),
the logarithm is bounded by an integrable multiple of
\(1+|\log y|\), while for large \(y\) the one-object subtraction
cancels the terms linear in the individual pole weights and gives
\(O(y^{-4})\). The remaining non-infrared contribution is
exponentially smaller by the exact factorisation and the bounds above.
Dominated convergence therefore yields
\begin{equation}
    E_{\rm int}
    =
    E_{\rm IR}
    +
    o(e^{-md/2}).
    \label{app:eq:energy-remainder}
\end{equation}

The same argument applies to localised objects of fixed finite size.
If \(I_i=\operatorname{supp}W_i\) are compact and
\(\ell=\operatorname{dist}(I_1,I_2)\), the reduced-resolvent bound
holds uniformly for \(x\in I_1\), \(y\in I_2\), so that
\[
    \left\|
        W_1^{1/2}G_\xi^\perp W_2^{1/2}
    \right\|
    =
    O(e^{-\mu\ell}).
\]
Since \(\ell=d+O(1)\) for objects of fixed size, the preceding
large-distance estimates are unchanged.

\subsection{Control of the planar momentum integral}
\label{app:planar-extension}

For the planar problem the Euclidean frequency is replaced by
\(p=\sqrt{\xi^2+\mathbf k_\parallel^2}\), and
\[
    G_p^\perp(x,y)
    =
    G_p(x,y)
    -
    \frac{\psi_0(x)\psi_0(y)}{p^2}.
\]
The infrared estimate
\eqref{app:eq:reduced-kernel-bound} is uniform on bounded intervals
of \(p\). At large \(p\), the ordinary Jost estimate together with
the exponential zero-mode tails gives, for every \(0<\mu<m\),
\begin{equation}
    \left|
        G_p^\perp(a_1,a_2)
    \right|
    \leq
    C_\mu
    \frac{e^{-\mu d}}{m^2+p^2}.
    \label{app:eq:global-reduced-bound}
\end{equation}
Consequently,
\begin{equation}
    \left|\log\Delta_\perp(p)\right|
    \leq
    C_\mu
    \frac{e^{-2\mu d}}{(m^2+p^2)^2},
    \label{app:eq:planar-Deltaperp-bound}
\end{equation}
and its radial integral with measure \(p^2dp\) is
\(O(e^{-2\mu d})\).

The pole-factor remainder is controlled as in~\ref{app:infrared-reduction}. With
\(p=\sqrt{s_d}\,y\), the measure becomes
\(p^2dp=s_d^{3/2}y^2dy\); the corresponding majorant behaves as
\(y^2|\log y|\) at the origin and as \(y^{-2}\) at infinity, and is
therefore integrable. The infrared reduction may thus be taken
through the radial momentum integral.

Since \(s_d=O(e^{-md})\), the leading infrared term is
\(O(e^{-3md/2})\). Choosing \(3m/4<\mu<m\) gives
\[
    e^{-2\mu d}
    =
    o(e^{-3md/2}),
\]
so direct propagation through the massive sector remains subleading
after the worldvolume momentum integration. This proves the remainder
estimate in Eq.~\eqref{eq:qft-planar-energy}.

\subsection{Microscopic sigma-model realisation:
classical and Gaussian consistency}
\label{app:sigma-realisation}

We substantiate the microscopic construction of
Subsec.~\ref{subsec:PT-microscopic}. Parametrising the
constrained field as
$\bm N=2(\sin(\phi/2)\cos\theta,
\sin(\phi/2)\sin\theta,\cos(\phi/2))$,
the sector $\theta=0$ obeys the sine-Gordon equation,
independently of $W$. The kink $\phi_K$ therefore remains
a classical solution with zero, one or both impurities:
the impurity term and its first variation vanish at $N_y=0$.
For any subset $J$ of repulsive impurities, completing
the square in the topological sector
$\bm N(-\infty)=(0,0,2)$,
$\bm N(+\infty)=(0,0,-2)$ gives
\begin{equation*}
\begin{aligned}
E_{\rm cl}={}&\frac8{g^2}
+\frac1{2g^2}\int\dd x\,
\left(\phi'-2\sin\frac\phi2\right)^2\\
&+\frac2{g^2}\int\dd x\,
\sin^2\frac\phi2\,(\theta')^2
+\frac1{2g^2}\sum_{i\in J}\lambda_iN_y^2(a_i)
\ \geq\ \frac8{g^2}.
\end{aligned}
\end{equation*}
The wall ($\theta=0$,
$\phi_K'=2\sin(\phi_K/2)$) saturates this bound
for every $J$. Thus all four classical energies coincide,
and $E_{\rm int}^{\rm cl}=0$ exactly: there is no classical
backreaction to neglect.

For the fluctuation expansion it is preferable to use
regular coordinates $y=N_y$ and $\phi$, with
$\bm N=(\sqrt{4-y^2}\sin(\phi/2),y,
\sqrt{4-y^2}\cos(\phi/2))$.
The constrained Lagrangian becomes
\begin{equation}
\begin{aligned}
g^2\mathcal L={}&
\frac12\left(1-\frac{y^2}{4}\right)(\partial\phi)^2
+\frac12\left(1-\frac{y^2}{4}\right)^{-1}
(\partial y)^2\\
&-(1-\cos\phi)
-\frac{y^2}{2}
\left[\cos^2\frac\phi2+W(x)\right].
\end{aligned}
\label{app:eq:sigma-chart}
\end{equation}
Setting $\phi=\phi_K+g\zeta$ and $y=g\eta$,
all linear terms vanish by the background equations.
Since the $y$-dependent terms are even, there is no
quadratic $\zeta$--$\eta$ mixing. On the static kink,
the kinetic contribution
$-y^2(\partial\phi)^2/8$ gives
$+(\phi_K')^2\eta^2/8$, yielding precisely
the transverse operator~\eqref{eq:PT-sigma-Hessian}.
The longitudinal operator is $K_\zeta=\Kpt$,
independently of $W$. The unperturbed angular zero
mode has profile $\psi_0=\sech x/\sqrt2$,
whereas the translational mode belongs to
the longitudinal sector.

Holding the wall centre fixed may be implemented by
the positive term
$E_{\rm pin}=\gamma N_z^2(0)/(2g^2)$, $\gamma>0$.
It vanishes on the centred wall, preserves axial
symmetry and shifts only
$K_\zeta\mapsto\Kpt+\gamma\delta(x)$.
This block is identical in the four configurations
entering the interaction subtraction and hence cancels
from their one-loop determinants. The surviving
transverse determinant contains the full P\"oschl--Teller
resolvent, including its massive continuum.

Finally, the internal orientation is a compact collective
coordinate $\vartheta$. For a rigidly rotated wall,
\begin{equation}
I=\frac8{g^2},\qquad
V_{\rm imp}(\vartheta)=
\frac2{g^2}\sum_i\lambda_i\sech^2a_i\,
\sin^2\vartheta.
\label{app:eq:sigma-rotor}
\end{equation}
The corresponding bare harmonic frequency is
$\Omega_{\rm bare}^2=
\frac12\sum_i\lambda_i\sech^2a_i$;
the massive-mode dressing is retained in the full
Gaussian determinant. Localisation of the angular
wavefunction requires $I\Omega_{\rm bare}\gg1$.
For fixed pair centre and large $d$, this entails
$g^2e^{d/2}\ll1$ for impurity strengths of order one; for weak
identical impurities, $\lambda\ll1$, it requires
$\lambda\gg g^{4}e^{d}$, up to $O(1)$ factors.
Thus the large-distance result has a joint semiclassical
domain; it is not asserted as an exact asymptotic
of the nonlinear sigma model at fixed $g$.

\end{document}